\documentstyle[12pt,epsf]{article}

\renewcommand{\thefootnote}{\fnsymbol{footnote}}
\newcommand{\newsection}{
\setcounter{equation}{0}
\section}

\def\appendix#1{
  \addtocounter{section}{1}
  \setcounter{equation}{0}
  \renewcommand{\thesection}{\Alph{section}}
  \section*{Appendix \thesection\protect\indent \parbox[t]{11.715cm} {#1}}
  \addcontentsline{toc}{section}{Appendix \thesection\ \ \ #1}
  }

\newcommand{\tr}[1]{\:{\rm tr}\,#1}

\def\e{{\,\rm e}\,}
\def\d{\partial }

\def\eop{\vspace*{\fill}\pagebreak}
\newcommand{\rf}[1]{(\ref{#1})}
\newcommand{\eq}[1]{Eq.~(\ref{#1})}
\def\be{\begin{equation}}
\def\ee{\end{equation}}
\def\beq{\begin{equation}}
\def\eeq{\end{equation}}
\def\bea{\begin{eqnarray}}
\def\eea{\end{eqnarray}}
\def\LB{\left (}
\def\RB{\right )}

\def\g{\gamma}
\def\cl{{\rm cl}}
\def\s{{\mbox{\boldmath $\sigma$}}}

\newcommand{\non}{\nonumber \\*}
\newcommand{\ie}{{\it i.e.}\ }

\newcommand{\ra}{\rightarrow}
\def\bl{\Bigl(}

\def\br{\Bigr)}

\begin{document}

\begin{titlepage}
\begin{flushright}
HIP--1997--14/TH\\
ITEP--TH--11/97\\
NBI--HE--97--13\\
hep-th/9704075\\
March, 1997
\end{flushright}
\vspace{1cm}

\begin{center}
{\LARGE Three Introductory Lectures in Helsinki\\[.4cm]
 on Matrix Models of Superstrings}\\
\vspace{1.4cm}
{\large Yuri Makeenko}\footnote{E--mail:
makeenko@vxitep.itep.ru \ \ \ \
 makeenko@nbi.dk \ } \\
\vskip 0.2 cm
{\it Institute of Theoretical and Experimental Physics,}
\\ {\it B. Cheremushkinskaya 25, 117259 Moscow, Russia}
\\ \vskip .1 cm
and  \\  \vskip .1 cm
{\it The Niels Bohr Institute,} \\
{\it Blegdamsvej 17, 2100 Copenhagen {\O}, Denmark}
\end{center}
\vskip 1.5 cm
\begin{abstract}
These are short notes of three introductory lectures on
recently proposed matrix models of Superstrings and M~theory
given at 5th Nordic Meeting on Supersymmetric Field and String
Theories in Helsinki (March 10--12, 1997).

\vspace{6pt}
Contents:
\begin{itemize}\vspace{-4pt}
\addtolength{\itemsep}{-4pt}
\item[1.] M(atrix) theory of BFSS,
\item[2.] From IIA to IIB with IKKT,
\item[3.] The NBI matrix model.
\end{itemize}
\end{abstract}

\end{titlepage}
\setcounter{page}{2}
\renewcommand{\thefootnote}{\arabic{footnote}}
\setcounter{footnote}{0}

\section*{Preface}

These lecture notes are based on three lectures at
5th Nordic Meeting on Supersymmetric Field and String
Theories in Helsinki (March 10--12, 1997).

The request from the Organizers was to make the lectures
understandable for graduate students. For this reason
the literal title of the transparencies was
\begin{center}
           {\bf A guide (for graduate students) }\\[.2mm]
           {\bf on how to read (and write) } \\[.2mm]
           {\bf papers on hep-th on} \\[.2mm]
           {\bf Matrix Models of Superstrings}
\end{center}
and the presentation was along this line.

The main goal of the lectures was to introduce the audience
into a fast developing subject of application
of matrix models in Superstring Theory. The knowledge
of superstrings is assumed at the level of about first
nine chapters of the book by Green, Schwartz and Witten.
A fascinating subject of string dualities and, correspondingly,
applications to M~theory is practically left outside for this reason.
No preliminary knowledge of matrix models is assumed.
The proper terminology, which is clarified in the lectures, is listed in
the next page.

Each of the three lectures is mostly concentrated around
one of the three selected papers~\cite{BFSS96,ikkt,FMOSZ97}.
The references in the
text are only to the results quoted. More complete list of
references can be found clicking a mouse on the number of
citations to the pioneering paper of
Banks, Fischler, Shenker and Susskind~\cite{BFSS96}
in HEP database at SLAC.

\begin{theindex}

  \item BPS, 6, 18
  \item branes
    \subitem anti-parallel, 16, 19
    \subitem moving, 20
    \subitem parallel, 16, 19
    \subitem rotated, 16, 21

  \indexspace

  \item compactification, 5

  \indexspace

  \item D(irichlet)-brane, 6
  \item D-instanton, 12
  \item D-particle, 6
  \item D-string, 12, 15
  \item duality, 5
    \subitem S
    \subitem T

  \indexspace

  \item effective action, 17
  \item eleven-dimensional, 5

  \indexspace

  \item Green--Schwartz, 12, 13, 24

  \indexspace

  \item large N, 5--25

  \indexspace

  \item matrix model, 4
  \item M(atrix) theory, 5
  \item membrane, 10, 11

  \indexspace

  \item Nambu-Goto action, 13, 24
  \item NBI action, 22

  \indexspace

\vspace{7cm}

  \item p-brane, 6, 16 

  \indexspace

  \item quantization
    \subitem 1st, 9
    \subitem 2nd, 14

  \indexspace

  \item Ramond-Ramond, 5
  \item reduced models
    \subitem Eguchi-Kawai, 12, 17

  \indexspace

  \item Schild action, 13
  \item string
    \subitem closed, 19
    \subitem open, 6, 19
  \item superstring
    \subitem IIA, 5, 12
    \subitem IIB, 12
  \item supersymmetry
    \subitem N=2, 13, 14
    \subitem target-space, 4
  \item super Yang-Mills, 12, 14

  \indexspace

  \item ten-dimensional, 5

  \indexspace

  \item Witten, 7

\end{theindex}

\newsection*{Introduction}
\setcounter{section}{0}

The standard non-perturbative approach to bosonic (Polyakov) string,
which is based on discretized random surfaces and matrix
models, exists since the middle of the eighties \cite{Kaz85,Dav85,ADF85}.
The main result of the investigations (both analytical and numerical) within
this approach is that the bosonic string is not in the stringy
phase but rather in a {\em branched polymer}\/ phase when the dimension $D$ of
the embedding space is larger than one%
\footnote{For a review see Ref.~\cite{Mak96}.}.
This is the way how the tachyonic problem is resolved for $D>1$.
In other words the perturbative vacuum with the tachyon is unstable
and the system chooses a stable vacuum which is {\em not}\/ associated
with strings.

A question immediately arises what about superstrings where
the GSO-projection kills tachyons (at least perturbatively). This is a
strong argument supporting the expectation for superstrings to live in
a stringy phase, which agrees with the common belief that fermions
smooth out the dynamics.

The attempts (not quite successful until very recent time)
of discretizing superstrings
are performed starting from~\cite{MS90}.
The problem resides, roughly speaking, in the fact that is not
easy to discretize the target-space supersymmetry (SUSY).
A progress had been achieved only for the simplest case
of pure two dimensional supergravity
which can be associated with a supereigenvalue model~\cite{AG92}.
For a more detail review, see Ref.~\cite{Mak96}.

The dramatic recent progress in a non-perturbative formulation of superstrings
by supersymmetric matrix models, which has occurred during last few months, is
the subject of these lecture notes. I shall mostly concentrate on
ten dimensional superstrings practically
leaving outside presumably most interesting
question of constructing the fundamental Lagrangian of eleven dimensional
M~theory in the language of matrix models.

\eop
\newsection{M(atrix) theory of BFSS~\protect{\cite{BFSS96}}}

Eleven dimensional M~theory combines different ten dimensional
superstring theories (IIA, IIB, \ldots), which are in fact related
by duality transformations, into a single fundamental theory.
BFSS proposed~\cite{BFSS96} to describe it by a supersymmetric
matrix quantum mechanics in the limit of infinite matrices.
This construction is called M(atrix) theory.

\subsection{The set up}

The point of interest of Ref.~\cite{BFSS96} is $D=10+1$ dimensional
M~theory (characterized by its Planck's length, $l_p$).
The eleven coordinates
$$
x^\mu=\left( t, x^i, x^{11}  \right)~~~~~~~~\left(i=1,\ldots,9\right)
$$
are split into time, $t$, the nine {\em transverse}\/ ones, $x^i$ or
$x^\perp$, and the {\em longitudinal}\/ one, called $x^{11}$, which
is compactified:
$$
x^{11}=x^{11}+2\pi R\,.
$$
The radius of compactification $R$ plays the role of an infrared cutoff
in the theory.

The system is considered in the Infinite Momentum Frame (IMF),
which is the same as the light cone frame, boosting along the
longitudinal axis. The same notations $t$ and $x^{11}$ are
used for $t\pm x^{11}$. The advantage of using IMF is that
only positive momenta $p_{11}$ are essential while systems with
zero or negative $p_{11}$ do not appear as independent dynamical
degrees of freedom. A price for this is the absence of
manifest Lorentz invariance.

Due to compactness all systems have (positive) longitudinal
momentum
\be
p_{11}=\frac NR ~~~~~~~~(N>0)\,,
\label{p11}
\ee
where $N>0$ is integer. At the end of calculations $R$ and $N/R$
should tend to infinity,
\be
R\ra\infty\,,~~~~~~~\frac NR \ra \infty \,,
\label{infinite}
\ee
to get {\em uncompactified}\/
infinite momentum limit of 11D theory.
$N$ will be identified in what follows
with the Ramond-Ramond (RR) charge of the system.

\subsection{10D versus 11D language}

M~theory with compactified $x^{11}$ is by construction
type {\em IIA superstring}\/ in $D=9+1$ dimensions.
The parameters $R$ and $l_p$ of eleven dimensional M~theory
and those $g_s$ and $l_s$ of the ten dimensional superstring
are related by
\be
R=g_s^{2/3} l_p\,,~~~~~~~~l_s=g_s^{-1/3} l_p\,,
\label{relation}
\ee
where $g_s$ is the {\em string coupling constant}\/ and
$l_s\equiv \sqrt{\alpha^\prime}$ is the string length scale
related to the {\em string tension}\/ $T$ by
\be
T=\frac{1}{2\pi\alpha^\prime} \,.
\ee
The 11D M~theory is in turn a strong coupling limit of 10D IIA
superstring, since~\rf{infinite} is guaranteed as $g_s\ra\infty$.

No perturbative string states carry RR charge $\Longrightarrow$
they are associated with vanishing momentum $p_{11}$.
1 unit of RR charge is carried by D0-brane of Polchinski~\cite{Pol95}
for which
\be
p_{11}=\frac 1R
\label{p11=1}
\ee
in accord with \eq{p11} at $N=1$.

The {\em low-energy}\/ limit of M~theory is 11D supergravity having
256 massless states: 44 gravitons, 84 three-forms and 128 gravitinos.
These 256 states are referred to as {\em supergravitons}\/ which are
{\em massless}\/ as 11D objects $\Longrightarrow$ they are
Bogomolny--Prasad--Sommerfield
(BPS) saturated states in 10D theory. Their 10D mass $\sim 1/R$.

States with $N\neq 1$ are not associated with elementary D0-branes.
The states with $N>1$ are bound composites of $N$ D0-branes as
is discussed in the next subsection.

\subsection{The appearance of matrices}

The world-volume of a p-brane is parametrized by $p+1$ coordinates
$\xi_0,\ldots,\xi_p$. The p-branes emerge as classical solutions in 10D
supergravities%
\footnote{For a review, see Ref.~\cite{Ste97}.}
which describe low-energy limits of 10D superstrings.
They possess an intrinsic abelian gauge field $A_\alpha\left(\xi\right)$
$(\alpha=0,\ldots,p)$ which can be viewed as tangent (to p-brane)
components of 10D abelian gauge field reduced to p-brane.
Otherwise, the remaining $9-p$ components of the 10D abelian gauge
field, which are orthogonal to the p-brane, are associated with its
coordinates~\cite{Pol95}
\be
X_i \left(\xi\right) = 2\pi \alpha' A_i\left(\xi\right)~~~~~~~
(i=p+1,\ldots,9) \,.
\ee

\begin{figure}[tp]
\hspace*{2cm}
\epsfysize=6cm \epsfbox{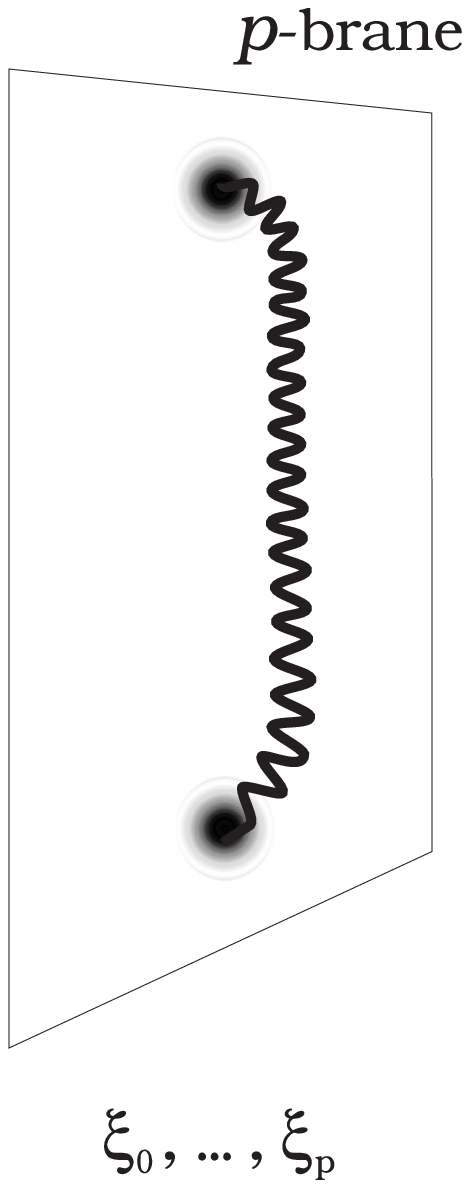}
\caption[x]
{\hspace{0.2cm}\parbox[t]{13cm}
{\small Dp-brane (depicted by a hyperplane parametrized by
the coordinates $\xi_0,\ldots,\xi_p$)
and the fundamental string.}}
   \label{fi:pbrane}
\end{figure}

A D(irichlet) p-brane can emit a fundamental open string which has
the Dirichlet boundary condition on a $p+1$ dimensional hyperplane
and the Neumann boundary condition in the $9-p$ dimensional
bulk of space. This string can end either on the same Dp-brane or
on another one as is illustrated by Fig.~\ref{fi:pbrane}.

\begin{figure}[tb]
\hspace*{2cm}
\epsfxsize=10cm \epsfbox{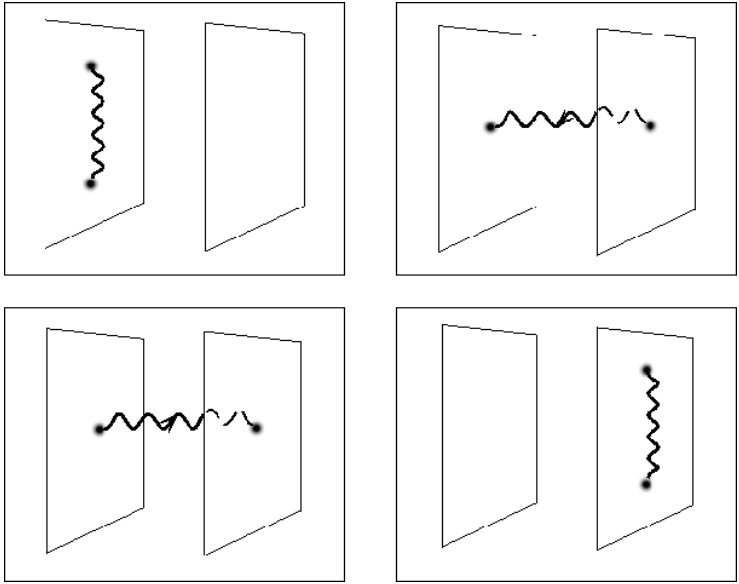}
\vspace*{1cm}
\caption[x]
{\hspace{0.2cm}\parbox[t]{13cm}
{\small
Appearance of matrices in the example of bound states
of two parallel D-branes \mbox{($N=2$).}
The fundamental string can
begin and end either at the same or different D-branes.
Since the string is oriented, there are four massless
vector states when the branes are practically on
the top of each other. They form a representation of U$(2)$.
}}
   \label{fi:wima}
\end{figure}

If one has $N$ parallel
Dp-branes separated by some distances in the $9-p$
dimensional space, then massless vector states emerge only when the string
begins and ends at the same brane, so the gauge group U$(1)^N$
appears in a natural way. Since the energy of strings
stretched between different D-branes is
\be
M\sim T\,| X^{i}- X^{j} | \,,
\ee
more massless vector states appear when the branes are practically on
the top of each other. Since the string is oriented, all possible massless
states when the string begins and ends either on same or different
Dp-branes form a U$(N)$ multiplet when strings are very short.
The example of $N=2$ is illustrated by Fig.~\ref{fi:wima}.
This is how hermitian $N\times N$
matrices appear in the description of
bound composites of $N$ Dp branes according to Witten~\cite{Wit95}.

For our case of $N$ D0-branes, their coordinates $X_i(t)$ become 9
Hermitean $N\times N$ matrices $X_i^{ab}\left(t\right)$
(accompanied by the fermionic superpartners
$\theta^{ab}_\alpha\left(t\right)$ which
are $16$ component nine dimensional spinors). They can be thought
as spatial components of the vector field in ten dimensional
super Yang--Mills theory after reduction to zero
space dimension (same for the superpartners).
$N$ is associated with the value of the RR charge of these states.

\subsection{The fundamental Lagrangian}

The possibility of formulating the fundamental Lagrangian of M~theory
as a matrix model is formulated in Ref.~\cite{BFSS96} as the
\begin{description}
\item[Conjecture:] M~theory in IMF is a theory with
the only dynamical degrees of freedom of D0-branes.
\end{description}
In other words {\em all}\/ systems are composed of D0-branes.
Therefore, the fundamental Lagrangian of M~theory is completely
expressed via the hermitian $N\times N$ matrices
$X_i^{ab}\left(t\right)$ describing coordinates of
D0-branes (and their fermionic superpartners $\theta^{ab}_\alpha(t)$), so that
$$
\fbox{ M~theory = M(atrix)~theory }\,.
$$

M(atrix) theory is described (in units of $l_s=1$) by the Lagrangian
\be
L={1\over 2g_s}\tr \left( \dot{ X^i}\dot{ X^i} + 2{\theta^T}
\dot{\theta} - {1\over 2}[X^i,X^j]^2 - 2{\theta^T} \g_i [\theta ,X^i
]\right).
\label{Lagrangian}
\ee
Here $N\ra\infty$ in order to satisfy~\rf{infinite}.

Changing the units to those where eleven dimensional $l_p=1$
and introducing
$$
Y={X}/{g^{1/3}_s}\,,
$$
\eq{Lagrangian} can be rewritten as
\be
L= \tr \left( {1 \over 2R} D_t{Y^i} D_t{Y^i}- {1\over 4} R\ [Y^i,Y^j]^2 -
{\theta^T}  D_t{\theta} - R\ {\theta^T}\g_i [\theta ,Y^i] \right),
\label{LagrangianY}
\ee
where
\be
D_t=\partial_t + i A_0
\ee
is the covariant derivative with respect to the $A_0$ field.
Equation~\rf{Lagrangian} is written in the $A_0=0$ gauge.

The Lagrangian~\rf{LagrangianY} is invariant under two {\em SUSY
transformations}
\bea
\delta_{\rm SUSY} X^i &=& - 2{\epsilon}^T \g^i \theta\,, \\
\delta_{\rm SUSY} \theta &=& \frac 12  \left( D_t X^i \g_i +\g_- +
\frac 12 [X^i ,X^j ]\; \g_{ij} \right) \epsilon + \epsilon^{\prime}\,, \\
\delta A_0 & =& - 2{\epsilon^T} \theta\,,
\eea
where $\epsilon$ and $\epsilon^\prime$ are two independent 16
component ($t$-independent) parameters. It is seen from this formula
that $A_0$ is needed to close the SUSY algebra.

\subsection{Matrix quantum mechanics}

The {\em Hamiltonian}\/ which is associated with the
Lagrangian~\rf{LagrangianY} reads
\be
H=R\tr \left\{{{{ \Pi_i \Pi_i}\over 2}+{1\over 4}\ [Y^i,Y^j]^2}+{\theta^T}
\g_i [\theta,Y^i]\right\},
\label{Hamiltonian}
\ee
where $\Pi_i$ is the canonical conjugate to $Y^i$.  As is usual for
fermions, a half of $\theta^{ab}_\alpha$ plays the role of coordinates and
the other half plays the role of canonical conjugate momenta
in the language of 1st quantization.

All finite energy states of the 10D
Hamiltonian~\rf{Hamiltonian} acquire infinite energy
as $R\rightarrow\infty$, \ie in the uncompactified 11D limit.
Only the states whose energy $\sim 1/ N$ as $N\ra\infty$ yield
\be
H\sim \frac RN = \frac 1{p_{11}}
\ee
as is expected since $p^2=2Ep_{11}-p_\perp^2$ in 11D IMF, so that
\be
p^2=0 \hbox{ \ (in 11D)}~~\Longrightarrow ~~E=\frac{p_\perp^2}{2p_{11}}
\hbox{ \ (in 10D)}
\label{E=}
\ee
in 10D.

The simplest states of the Hamiltonian~\rf{Hamiltonian} is
when the matrices $Y^i$ are diagonal with only one nonvanishing
diagonal component and all $\theta$'s equal zero.
For nonvanishing $p_\perp$ --- the eigenvalue of
$\Pi_\perp$ --- \eq{Hamiltonian} yields
\be
E=\frac R2 p_\perp^2 =\frac{p_\perp^2}{2p_{11}}
\label{1E=}
\ee
since the commutators vanish. Thus we get \eq{p11=1} with $N=1$ and this
state corresponds to a single D0-brane in 10D language.

Each of these states is accompanied by the fermionic superpartners
and they form a representation of the algebra of 16 $\theta$'s
with
$$
2^{16/2} = 2^8 = 256
$$
components. They are exactly 256 states of supergraviton in 11D.
In the 10D language these are BPS states of the mass $\sim 1/R$
which become massless in the uncompactified limit $R\ra\infty$.

A more general eigenstate of the Hamiltonian~\rf{Hamiltonian}
has a form of the {\em diagonal}\/ $N\times N$ matrix
\be
Y_i = \left(
\begin{array}{ccc}
Y_i^{(1)} & & \\
 & \ddots & \\
 & &  Y_i^{(N)}
\end{array}
\right).
\label{diagonal}
\ee
The commutator obviously vanishes in this case.

It is convenient to split the U$(N)$ group as
\mbox{U$(1)\otimes$ SU$(N)$} and to associate the U$(1)$ part with
the center mass coordinate
\be
Y_i(cm)=\frac 1N \tr Y_i \,.
\ee
Then
\be
p_i(cm)= \tr \Pi_i = \frac NR \dot{Y}_i (cm)\,,
\ee
and using $p_{11}=N/R$ we get the usual relation
\be
\frac{1}{p_{11}} p_i (cm) = \dot{Y}_i (cm)
\ee
between transverse velocity and momentum.

Interaction states are described in this construction by
{\em non-diagonal}\/ matrices. They correspond to
{\em scattering states}\/ of supergravitons in 11D.
The interaction of supergravitons at the tree level
is correctly reproduced within M(atrix) theory.

\subsection{The relation to membranes\label{Weyl}}

The Hamiltonian~\rf{Hamiltonian} of the $N\ra\infty$
supersymmetric quantum mechanics looks pretty much like
the one~\cite{dWHN88} for a 11D supermembrane in IMF.
While there are no truly stable finite energy membranes in
the decompactified limit, there exist very long lived
classical membranes.

The membrane action can be derived in the Weyl basis on gl$(N)$,
which is given by two unitary $N\times N$ matrices $g$ and $h$
(clock and shift operators) obeying
\bea
&hg=\omega gh\,,~~~\omega=\e^{2\pi i/N}\,,& \\
&h^N=1=g^N\,. &
\eea
Any hermitian $N\times N$ matrix $Z$ can be expanded in this
basis as
\be
Z=\sum_{n,m=1}^N Z_{n,m} g^m h^n \,.
\ee

As $N\ra\infty$, we can introduce a pair of canonical variables
$q$ and $p$, so that
\bea
g=\e^{ip}\,,~& &~h=\e^{iq}\,, \\
\left[ q \,,\, p\right]&=&\frac{2\pi i}{N} \,.
\label{canonical}
\eea
As usual in quantum mechanics, the last equality is possible
only as $N\ra\infty$.
Then, we have
\bea
\tr Z &\Rightarrow &N \int dp dq Z(p,q) \,,\\
\left[ X\,,\, Y\right] &\Rightarrow & \frac iN
\left\{  \partial_q X \partial_p Y
- \partial_p X\partial_q Y\right\}
\eea
for the trace and the commutator, and finally~\cite{BFSS96}
$$
\fbox{ M(atrix) action $\Longrightarrow$ Supermembrane action }
$$
as $N\ra\infty$.

A special comment is needed concerning the continuum spectrum
of the supermembrane~\cite{dWLN89}. From the point of view of
the M(atrix) theory, it is as a doctor ordered for describing
the supergraviton scattering states. The conjecture of
M(atrix) theory is that there exists a normalizable bound state
at the beginning of the continuum spectrum at $p^2=0$.

The emergence of membranes in M(atrix) theory can be seen from the
classical equations of motion
\be
\left[ Y^i\,,\left[ Y^j\,,\, Y^i \right] \right] =0\,,
~~~~ \left[Y^i \,, \,(\g_i\theta)_\alpha\right]=0
\label{Mce}
\ee
which are satisfied by static configurations.

An infinite membrane stretched out in the 8,9 plane
is given by~\cite{BFSS96}
\bea
Y^8=R_8 \sqrt{N} p\,,~~
Y^9=R_9 \sqrt{N} q\,, \non
\hbox{all other $Y$'s and $\theta$'s }=0 \,,~~~~~
\label{D-membra}
\eea
where $p$ and $q$ are $N=\infty$ matrices (operators), and
$R_8$ and $R_9$ are (large enough) compactification radii.
Equations~\rf{Mce} are satisfied by~\rf{D-membra} because
\be
\left[ Y^8\,, \, Y^9 \right] =\;\hbox{c-number}\,.
\ee
The membrane in this picture is built out of infinitely
many D0-branes.

The interaction between these membrane configurations
has been studied~\cite{AB96,LM96,Lif96} and compared with the
superstring results.

\eop
\newsection{From IIA to IIB with IKKT~\protect{\cite{ikkt}}}

M(atrix) theory naturally describes ten dimensional IIA superstring.
IKKT proposed~\cite{ikkt} another matrix model associated with IIB
superstring, which is in spirit of the Eguchi--Kawai large-N reduced ten
dimensional super Yang--Mills theory. This non-perturbative
formulation of IIB superstring is called the IKKT matrix model.

\subsection{Preliminaries}

IIB superstring differs from IIA superstring by chiralities of
the fermionic superpartners. They are opposite for IIA
superstring and same for IIB superstring.

As a consequence of this, Dp-branes of even $p$
($p=0,2,4,\ldots$)
are consistently incorporated by type IIA superstring theory while
type IIB superstring is associated with Dp-branes of {\em odd}\/ $p$
($p=-1,1,3,5,\ldots$)~\cite{Pol95}.
This is due to the rank of the antisymmetric field which is
odd for IIA superstring and even for IIB superstring.
Correspondingly, the analog of D0-brane (associated with $p=0$
in the IIA case) is D-instanton (associated with $p=-1$ in the IIB
case) and the analog of D-membrane (associated with $p=2$ in the IIA
case) is D-string (associated with $p=1$ in the IIB case).

In analogy with Ref.~\cite{BFSS96} where the fundamental Lagrangian
is expressed in terms of D0-branes, one might expect that IIB
superstring is described in terms of D-instanton variables, {\it i.e.}\
by the ten dimensional super Yang--Mills dimensionally reduced to a
point~\cite{Wit95}.

\subsection{Schild formulation of IIB superstring\label{Schi}}

The starting point in the IKKT approach is the Green--Schwartz
action of type IIB superstring theory with fixed $\kappa$-symmetry:
\be
S_{\rm GS}=-T \int d^2 \sigma \left\{
\sqrt{-\s^2}+2i\varepsilon^{ab} \d_a X^\mu \bar{\Psi}\g_\mu \d_b \Psi
\right\},
\label{GSaction}
\ee
where
\be
  \s^{\mu\nu} =\varepsilon^{ab} \d_a X^\mu \d_b X^\nu \,,
\ee
the vector index $\mu$ of $X^\mu\left(\sigma_1,\sigma_2\right)$
runs from $0$ to $9$ and the spinor
index $\alpha$  of $\Psi_\alpha\left(\sigma_1,\sigma_2\right)$ runs
from $1$ to $32$. The fermion $\Psi$ is a Majorana--Weyl
spinor in 10D which satisfies the condition $\g_{11}\Psi=\Psi$,
so that only $16$ components effectively remain.

The action~\rf{GSaction} is invariant under the ${\cal N}=2$
{\em SUSY transformation}
\bea
\delta_{\rm SUSY} \Psi_\alpha &=&
\frac{1}{2\sqrt{-\frac12 \s^2}} \s^{\mu\nu} \left(\g_{\mu\nu}
\epsilon\right)_\alpha +\xi_\alpha \,,\non
\delta_{\rm SUSY} X^\mu &=& 4i \bar\epsilon \g^\mu \Psi
\eea
whose parameters $\epsilon$ and $\xi$ do not depend on
$\sigma_1$ and $\sigma_2$.

The action~\rf{GSaction} can be rewritten in the Schild form
\be
S_{\rm Schild}=
\int d^2\sigma \left\{\sqrt{g}\,\alpha\, \Big(
\frac{1}{4}\{X^{\mu},X^{\nu}\}^2
-\frac{i}{2}\bar{\Psi}\g^{\mu}\{X_{\mu},\Psi\}\Big)
+\beta \sqrt{g}\right\},
\label{Saction}
\ee
where $\sqrt{g\left(\sigma_1,\sigma_2\right)}$
is positive definite scalar density (which is considered as an independent
dynamical variable) and the Poisson bracket is defined by
\beq
\{X,Y\} \equiv  \frac{1}{\sqrt{g}}\varepsilon^{ab}\partial_a X
\partial_b Y\,.
\label{PB}
\eeq
Note that $\sqrt{g}$ cancels in the fermionic term in the action.

The equivalence of~\rf{GSaction} and \rf{Saction} at
the classical level can be proven by using the classical equation
of motion for $\sqrt{g}$. Varying the Schild action~\rf{Saction}
with respect to $\sqrt{g}$, we get
\be
-\frac{1}{4} \alpha \frac{1}{\left(\sqrt{g}\right)^2}
\left(\varepsilon^{ab} \d_a X^\mu \d_b X^\nu \right)^2 +\beta =0\,.
\ee
Substitution of the solution
\be
\sqrt{g} = \frac 12 \sqrt{\frac{\alpha}{\beta}}
\sqrt{\left(\varepsilon^{ab} \d_a X^\mu \d_b X^\nu \right)^2 }
\ee
into~\rf{Saction} restores the Nambu--Goto form~\rf{GSaction}
of the Green--Schwartz action:
\be
S_{\rm NG} =T \int d^2 \sigma \left\{
\sqrt{\alpha \beta}
\sqrt{\left(\varepsilon^{ab} \d_a X^\mu \d_b X^\nu \right)^2 }
- \frac{i}{2}\, \alpha\,
\varepsilon^{ab} \d_a X^\mu \bar{\Psi}\g_\mu \d_b \Psi
\right\}.
\label{NGaction}
\ee

The action~\rf{NGaction} is invariant under the ${\cal N}=2$
SUSY transformation
\bea
\delta_{\rm SUSY} \Psi_\alpha &=& -\frac{1}{2} \sqrt{g}
   \{ X_\mu,X_\nu \}\left(\g^{\mu\nu}\epsilon\right)_\alpha
   + \xi_\alpha ,\non
\delta_{\rm SUSY} X^{\mu} &=& i\bar{\epsilon}\g^{\mu}\Psi \,,
\label{Ssym}
\eea
where the parameters $\epsilon$ and
$\xi$ do not depend again on $\sigma_1$ and $\sigma_2$.

Finally the {\em partition function}\/ in the Schild formulation of
IIB superstring is defined by the
path integral over the positive definite function
$\sqrt{g}$, and over $X^{\mu}$ and $\Psi_\alpha$:
\beq
 Z_{\rm Schild} = \int\, D \sqrt{g}\, D X^\mu\,
 D \Psi_\alpha \,\e^{- S_{\rm Schild}}\,.
\label{Smain}
\eeq
It is invariant under the SUSY transformation~\rf{Ssym}
since both the action~\rf{Saction} and the measure
$ DX^\mu D\Psi_\alpha$ are invariant.

Equations~\rf{Saction} and \rf{Smain} represent
IIB superstring in the Schild formalism with fixed
$\kappa$-symmetry~\cite{ikkt}.

In addition to the ${\cal N}=2$ SUSY transformation~\rf{Ssym},
the partition function is invariant at fixed $\sqrt{g}$ under
area-preserving or symplectic {\em diffeomorphisms}
\be
\delta_{\rm sdiff} X^\mu =  \{ X^\mu, \Omega\}\,,~~~~~
\delta_{\rm sdiff} \Psi_\alpha =  \{ \Psi_\alpha, \Omega\}
\label{Sdiff}
\ee
which is only a part of the whole
reparametrization (or diffeomorphism) transformations.
The invariance of the string theory under the whole group of
reparametrizations is restored when $\sqrt{g}$ is
transformed. The symmetry~\rf{Sdiff} reminds the non-abelian
gauge symmetry in Yang--Mills theory and is to be fixed for doing
perturbative calculations.

\subsection{The IKKT matrix model}

The IKKT matrix model can be obtained from the representation~\rf{Smain}
of IIB superstring in the Schild formalism by replacing
\bea
X_\mu\left(\sigma_1,\sigma_2\right) &\Longrightarrow& A_\mu^{ab} \,,
\label{r1} \\*
\Psi_\alpha\left(\sigma_1,\sigma_2\right) &\Longrightarrow&
\psi_\alpha^{ab} \,,
\label{r2}
\eea
where  $A_\mu^{ab} $ and $\psi_\alpha^{ab} $ are hermitian $n\times n$
bosonic and fermionic matrices, respectively.

The IKKT matrix model is defined by the partition function
\beq
Z= \sum_{n=1}^{\infty} \int dA_\mu \,d\psi_\alpha \, \e^{-S} \,,
\label{main}
\eeq
which is of the type of 2nd quantized (euclidean) field theory,
with the action
\beq
S=\alpha\left( -\frac{1}{4}\tr[A_{\mu},A_{\nu}]^2
            -\frac{1}{2}\tr (\bar{\psi}
           \g^{\mu}[A_{\mu},\psi])\right)+\beta n\, .
\label{action}
\eeq
The summation over the matrix size $n$ in \eq{main} implies
that $n$ is a dynamical variable (an analog of $\sqrt{g}$
in \eq{Smain}).

The action~\rf{action} and the measure $dA_\mu \,d\psi_\alpha $
in~\rf{main} are invariant under
the ${\cal N}=2$ SUSY transformation
\bea
\delta_{\rm SUSY}\psi^{ab}_\alpha &=& \frac{i}{2}
   [A_{\mu},A_{\nu}]^{ab}(\g^{\mu\nu}\epsilon)_\alpha
   +\xi_\alpha \delta^{ab}\,,\non
\delta_{\rm SUSY} A_{\mu}^{ab} &=& i\bar{\epsilon}\g_{\mu}\psi^{ab} \,,
\label{sym}
\eea
where the parameters $\epsilon$ and $\xi$ are numbers rather than
matrices, as well as under the SU$(n)$ gauge transformation
\bea
\delta_{{\rm gauge}} A_{\mu} &=& i \left[A_\mu,\omega\right], \non
\delta_{{\rm gauge}} \psi_{\alpha} &=& i \left[\psi_\alpha,\omega\right].
\label{gauge}
\eea

The formulas~\rf{sym} and \rf{gauge}
look like as if ten dimensional super Yang--Mills
theory is reduced to a point. For instance only the commutator
is left in the non-abelian field strength
\be
f_{\mu\nu} = i \left[A_\mu\,,\, A_\nu \right]
\label{fmunu}
\ee
and there are no space-time derivatives. However, the action~\rf{action}
coincides with the one of 10D super Yang--Mills dimensionally
reduced to zero dimensions only if $\beta=0$ and $n$ is fixed.
This differs the IKKT matrix model from a pure D-instanton matrix model.


As was argued in~\cite{ikkt},
if large values of $n$ and {\em smooth}\/ matrices $A_{\mu}^{ab}$ and
$\psi_{\alpha}^{ab}$ dominate in~\rf{main},
one substitutes
\bea
\left[\,\cdot\,,\,\cdot\,\right] & \Longrightarrow &
i \{\,\cdot\,,\,\cdot\,\}
\label{cb} \\*
\tr \ldots &\Longrightarrow& \int d^2 \sigma \sqrt{g} \ldots
\label{ti}
\eea
similarly to what is discussed in Subsect.~\ref{Weyl}.
Then the formulas~\rf{main} to \rf{gauge} for the IKKT matrix models
reproduce the ones~\rf{Saction} to \rf{Sdiff} for the Schild
formulation of IIB superstring.

This passage from the IKKT matrix model to the Schild
formulation of IIB superstring
can be formalized introducing the matrix function
\be
L\left(\sigma_1,\sigma_2\right)^{ab}= \sum_{m_1,m_2}
j_{m_1,m_2}\left(\sigma_1,\sigma_2\right) J_{m_1,m_2}^{ab}\,,
\label{defL}
\ee
where $J_{m_1,m_2}^{ab}$ form a basis for gl$(\infty)$ and
$j_{m_1,m_2}\left(\sigma_1,\sigma_2\right)$
form a basis in the space of functions of $\sigma_1$ and $\sigma_2$.
An explicit form of $j$'s depends on the topology of
the $\sigma$-space.
Explicit formulas are available for a sphere
and a torus.

With the aid of~\rf{defL} we can relate
matrices with functions of $\sigma_1$ and $\sigma_2$ by
\bea
A_\mu&=& \int d^2 \sigma \sqrt{g}\, X_\mu \,L \,,
\label{AvsX} \\*
X_\mu&=&\tr A_\mu\, L \,.
\label{XvsA}
\eea
These formulas result for smooth configurations
in Eqs.~\rf{cb} and \rf{ti}.
The word ``smooth'' means that
configurations can be reduced by a gauge transformation
to the form where high modes
are not essential in the expansions~\rf{AvsX} or \rf{XvsA}.

The commutators of $J$'s coincide with
the Poisson brackets of $j$'s as $n\ra\infty$.
This demonstrates the equivalence between
the group of symplectic diffeomorphisms
and the gauge group SU$(\infty)$ for smooth configurations.

\subsection{D-strings as classical solutions}

The classical equations of motion for the Schild action~\rf{Saction}
read
\be
\left\{ X^\mu\,,\left\{ X_\mu\,,\,X_\nu \right\}\right\}=0\,,
~~~~~\left\{ X^\mu\,,\left( \g_\mu\Psi\right)_\alpha \right\}=0 \,.
\label{Sce}
\ee
Their matrix model counterparts are
\be
\left[ A^{\mu},\,\left[ A_\mu,A_\nu \right]\right]=0\,,~~~~~
\left[ A^\mu\,, (\g_\mu\psi)_\alpha \right] =0\,,
\label{ce}
\ee
which are to be solved for $n\times n$ matrices ${A}_\mu$
at infinite $n$.

Since Eqs.~\rf{ce} look like \eq{Mce} for M(atrix) theory,
they possess {\em operator-like}\/ solutions of the
form~\rf{D-membra},
which are now associated with D-strings~\cite{ikkt}.
The solution associated with static D-string along 1st axis reads
\be
A_\mu^\cl = \left(\frac{T}{2\pi}q,\frac{L}{2\pi}p,0,\ldots,0   \right),
~~~~~\psi_\alpha^\cl=0\,,
\label{D-string}
\ee
where the (infinite) $n\times n$ matrices $p$ and $q$ obey
the canonical commutation relation~\rf{canonical}, while
$T/2\pi$ and $L/2\pi$ are (large enough) compactification radii.

The arguments in favor of identification of the classical
solution~\rf{D-string} with static D-string are
\begin{itemize} \vspace{-6pt}
\addtolength{\itemsep}{-6pt}
\item It is one dimension less than D-membrane of~\cite{BFSS96};
\item Interaction between the two D-strings is reproduced at large
       distances~\cite{ikkt};
\item It is a BPS state (a proper central charge of SUSY algebra
       exists~\cite{bss,CMZ97});
\item It can be extended to $p=3,5$~\cite{CMZ97,FS97}.
 \vspace{-6pt}
\end{itemize}

\subsection{Zoo of Dp-branes}

A solution associated with two D-strings has a block-diagonal form
and is built out of the ones given by \eq{D-string} for single D-strings.

The solution for two {\em parallel}\/ static D-strings
separated by the distance $b$ along 2nd axis reads~\cite{ikkt}
\begin{eqnarray}\label{backgr}
&A_{0}^\cl=\left(
 \begin{array}{cc}
 Q & 0 \\
 0 & Q \\
 \end{array}
 \right),~~~
 A_{1}^\cl=\left(
 \begin{array}{cc}
 P & 0 \\
 0 & P \\
 \end{array}
 \right),~~~
 A_{2}^\cl=\left(
 \begin{array}{cc}
 b/2 & 0 \\
 0 & -b/2 \\
 \end{array}
 \right),&\non
 &A_{3 }^\cl=\ldots=A_{9 }^\cl= 0\,, &
 \end{eqnarray}
where we have denoted
\be
Q\equiv\frac{T}{2\pi}q\,,~~~~~P\equiv\frac{L}{2\pi}p\,.
\label{PQ}
\ee

The solution associated with two {\em anti-parallel}\/ static
D-strings separated by the distance $b$ along 2nd axis is
\begin{eqnarray}\label{backgra}
&A_{0}^\cl=\left(
 \begin{array}{cc}
 Q & 0 \\
 0 & Q \\
 \end{array}
 \right),~~~
 A_{1}^\cl=\left(
 \begin{array}{cc}
 P & 0 \\
 0 & -P \\
 \end{array}
 \right),~~~
 A_{2}^\cl=\left(
 \begin{array}{cc}
 b/2 & 0 \\
 0 & -b/2 \\
 \end{array}
 \right),&\non
 &A_{3 }^\cl=\ldots=A_{9 }^\cl= 0\,. &
 \end{eqnarray}

The solution associated with two static D-strings
{\em rotated}\/ through the angle $\theta$ in the 1,2 plane
and separated by the distance $b$ along 3rd axis is
\begin{eqnarray}\label{backgrr}
&A_{0}^\cl=\left(
 \begin{array}{cc}
 Q & 0 \\
 0 & Q \\
 \end{array}
 \right),~~~
 A_{1}^\cl=\left(
 \begin{array}{cc}
 P & 0 \\
 0 & P\cos{\theta} \\
 \end{array}
 \right),~~~
 A_{2}^\cl=\left(
 \begin{array}{cc}
 P & 0 \\
 0 & P\sin{\theta} \\
 \end{array}
 \right), &\non
 &A_{3}^\cl=\left(
 \begin{array}{cc}
 b/2 & 0 \\
 0 & -b/2 \\
 \end{array}
 \right),~~~~
 A_{4 }^\cl=\ldots=A_{9 }^\cl= 0\,. &
 \end{eqnarray}

 The solution associated with one Dp-brane,
which extends~\rf{D-string} to $p>1$, is given by
\be
A_\mu^\cl = \left(P_1,Q_1,\ldots,P_{\frac{p+1}{2}},
Q_{\frac{p+1}{2}},0,\ldots,0   \right),
~~~~~\psi_\alpha^\cl=0\,,
\label{Dp}
\ee
where $P$'s and $Q$'s form $(p+1)/2$ pairs of operators (infinite matrices)
as in \eq{PQ} obeying canonical commutation relation on a torus
associated with compactification (of large enough radii $L_a/2\pi$)
along the axes $0,\ldots,p$, so that
\be
\omega_k=\frac{L_{2k-2}L_{2k-1}}{2\pi n^{\frac{2}{p+1}}}
~~~~~~~\left(k=1,\ldots,\frac{p+1}{2} \right)
\label{omega}
\ee
is fixed as $n\ra\infty$.
This is because of the fact that the full Hilbert space
of the dimension $n$ is represented as the tensor product
of $(p+1)/2$ Hilbert spaces of the dimension $n^{\frac{2}{p+1}}$
each~\cite{bss}. The value of $n$ is related
to the $p+1$ dimensional volume
\be
V_{p+1} \equiv L_0\,L_1\cdots L_p
\label{volume}
\ee
of the p-brane by
\be
n=V_{p+1} \prod_{i=1}^{\frac{p+1}{2}} (2\pi \omega_i)^{-1}\,.
\ee
These formulas allows one to extract world-volume characteristics of
Dp-branes from the matrix model.

A general multi-brane solution has a block-diagonal form and is built
out of single p-brane solutions~\rf{Dp} quite similar to
\rf{backgr}--\rf{backgrr}.

\subsection{One-loop effective action}

The calculation of the one-loop effective action in the
IKKT matrix model at fixed $n$ can be performed for
an arbitrary background, $A_\mu^{\rm cl}$ and $ \psi^{\rm
cl}_\alpha=0$, obeying the classical equations of motion~\rf{ce}.
The calculation is quite similar to the one in the Eguchi--Kawai
reduced model.

Expanding around the classical solution
\be
A_\mu = A_\mu^{\rm cl} + a_\mu
\ee
and adding the gauge fixing and ghost terms to the action~\rf{action}:
\be
S_{\rm g.f.}= -\tr \left( \frac 12
\left[ A_\mu^{\rm cl} \,,\, a_\mu\right]^2 +
\left[ A_\mu^{\rm cl} \,,\, b\right]
\left[ A_\mu^{\rm cl} \,,\, c\right]
\right),
\ee
where the matrices $b$ and $c$ represent ghosts,
we get~\cite{ikkt}
\be\label{seff}
W=\frac{1}{2}{\rm Tr}\ln(P^2\delta_{\mu
\nu}-2iF_{\mu\nu})-\frac{1}{4}{\rm Tr}\ln\LB \Big(
P^2+\frac{i}{2}F_{\mu\nu}\g^{\mu\nu}\Big)\LB\frac{1+\g_{11}}{2}
\RB\RB
-{\rm Tr}\ln P^2\,.
\ee
Here the {\em adjoint}\/
operators $P_{\mu}$ and $F_{\mu\nu}$ are defined
on the space of matrices by
\be
P_{\mu}=\left[A_{\mu }^\cl,\cdot\,\right],~~~~~
 F_{\mu\nu}=\left[f_{\mu \nu }^\cl,\cdot\,\right]=
i\left[\left[A_\mu^\cl,A_\nu^\cl\right],\cdot\,\right] .
\label{PF}
\ee
For the solution~\rf{Dp},
${\rm Im}\:W$ vanishes for $p=1,3,5,7$ since
$P_{\mu}=0$ at least in one direction.

The first term on the right hand side of \eq{seff} comes from
the quantum fluctuations of $A_\mu$, the second and third terms
which come from fermions and ghosts have the minus
sign for this reason. The extra factor $1/2$ in the first
and second terms is because
the matrices $A$ and $\psi$ are hermitian.

If $A_\mu^{\rm cl} $ is diagonal
\be
{A}_\mu^\cl= \hbox{diag}\left(p_\mu^{(1)}, \ldots,
p_\mu^{(n)} \right),~~~~~\psi^{\rm cl}_\alpha=0\,,
\label{D-instanton}
\ee
which is a solution of \eq{ce} associated with the flat space-time,
then $F_{\mu\nu}=0$ and
\be
W=\left( \frac 12 \cdot 10 -\frac 14 \cdot 16 -1 \right)
{\rm Tr}\ln P^2=0\,.
\ee
The plane vacuum is a BPS state.

The same is true (to all loops)
for any ${A}_\mu^\cl$ whose commutator is diagonal:
\be
\left[A_\mu^\cl,A_\nu^\cl\right] = c_{\mu\nu} {\bf 1}_n \,,
\label{BPS}
\ee
where $c_{\mu\nu}$ are c-numbers
rather than matrices. Such solutions
preserve~\cite{ikkt,bss} a half of SUSY and are BPS states.
The solution~\rf{backgr} associated with parallel D-strings
is an example of such a BPS state.

For a general background ${A}_\mu^\cl$, the matrix $F_{\mu\nu}$ can
always be represented in the canonical (Jordan) form
 \begin{equation}\label{stc}
 F_{\mu\nu}=\left(
 \begin{array}{ccccc}
 0 & -\omega_1 & & & \\
 \omega_1 &0& & & \\
  & &\ddots & & \\
 &&&0 & -\omega_5  \\
 &&&\omega_5 &0 \\
 \end{array}
 \right),
 \end{equation}
so that
\be
{\rm Tr}\ln(P^2\delta_{\mu
\nu}-2iF_{\mu\nu}) =\sum_{i=1}^5 {\rm Tr} \ln
((P^2)^2-4\omega_i^2)
\label{bos}
\ee
and
\be{\rm Tr}\ln\LB \Big(
P^2+\frac{i}{2}F_{\mu\nu}\g^{\mu\nu}\Big)\LB\frac{1+\g_{11}}{2}
\RB\RB=
\sum_{
 \begin{array}{c}
 \scriptstyle
 s_1,\ldots,s_5=\pm 1\\[-2.5mm]
 \scriptstyle
 s_1\ldots s_5=1
 \end{array}
 }{\rm Tr}\ln\left(P^2-\sum\limits_{i}\omega _i s_i\right)\,.
\label{fer}
\ee
There are 16 terms on the right hand side of \eq{fer} representing
the trace over $\g$-matrices. Equations~\rf{bos} and \rf{fer}
are most useful in practical calculations for the background of
Dp-brane given by~\rf{Dp}, when only $(p+1)/2$ of 5 omegas are nonvanishing.

\subsection{Brane-brane interaction}

The interaction between two Dp-branes is calculated by substituting the
proper classical solutions into~\rf{seff} and using Eqs.~\rf{bos},
\rf{fer}.

For parallel Dp-branes, $W=0$ is accordance with the general arguments
of the previous subsection.

For anti-parallel Dp-branes, we get (\cite{ikkt} for $p=1$,
\cite{CMZ97} for $p\geq 3$)
\be\label{otvet}
 W=-2n\int\limits_{0}^{\infty }\frac{ds}{s}\,\e^{-b^2s}
 \left[\sum_{i=1}^{\frac{p+1}{2}}
 \bl\cosh 4\omega _is-1\br-4\bl\prod_{i=1}^{\frac{p+1}{2}}\cosh
 2\omega _is-1\br\right]\prod_{i=1}^{\frac{p+1}{2}}
 \frac{1}{2\sinh 2\omega _is}.
 \ee
The asymptotics of this formula at large $b$:
 \be\label{1/b}
 W=-\,\frac{1}{16}\,n\left(\frac{5-p}{2}\right)!\,
 \left[2\sum_{i=1}^{\frac{p+1}{2}}\omega _i^4-
 \bl\sum_{i=1}^{\frac{p+1}{2}}\omega _i^2\br^2\right]
 \prod_{i=1}^{\frac{p+1}{2}}\omega _i^{-1}\left(\frac{2}{b}\right)^{7-p}
 +O\left(\frac{1}{b^{9-p}}\right),
\ee
agrees with the superstring calculation at large distances.

However, the superstring result~\cite{Pol95,Pol96,GG96} for the interaction
between two anti-parallel Dp-branes at arbitrary distances $b$,
which is given by the annulus diagram in the open-string language
or by the cylinder diagram in the closed-string language,
\be
W=-V_{p+1}\int_0^\infty \frac{dt}{t}
\frac{1}{\left(8\pi^2 \alpha' t\right)^\frac{p+1}{2}}
\e^{-{b^2 t}/{2\pi\alpha'}}
q^{-1}\frac{\prod_{n=1}^\infty (1-q^{2n-1})^8}
{\prod_{n=1}^\infty (1-q^{2n})^8}
\label{superstring}
\ee
with $q=\e^{-\pi t}$, does {\em not}\/ coincide with
the matrix-model result~\rf{otvet}.
There is no agreement even if one truncates to the lightest open string
modes.

A way out could be
to interpret~\cite{Tse97} the classical solutions
in the IKKT matrix model as D-branes with magnetic field,
in analogy with previous work~\cite{LM96}
on M(atrix) theory~\cite{BFSS96}.
An alternative possibility is to modify the
IKKT matrix model to better reproduce
the superstring calculation.

\eop
\newsection{The NBI matrix model~\protect{\cite{FMOSZ97}}}

Calculations of the brane-brane interaction in the matrix model can be
extended to the case of moving and rotated static Dp-branes.
The results agree with the superstring calculations for empty branes only at
large distances between them. This was one of the motivations of
Ref.~\cite{FMOSZ97} to modify the IKKT matrix model
introducing (instead of $n$)
an additional dynamical variable --- a positive definite
hermitian matrix $Y^{ab}$ --- which is the direct analog of $\sqrt{g}$ in
the Schild formulation of IIB superstring.
Integration over $Y^{ab}$ results in the Non-abelian
Born--Infeld (NBI) action which reproduces the Nambu--Goto version
of the Green--Schwarz action of IIB superstring.

\subsection{Parallel moving branes}

The operator-like solution to Eqs.~\rf{ce}, which is associated with
two parallel branes separated by the distance $b$ along the
$(p$+$2)$-th axis and
{\em moving}\/ with velocities $v$ and $-v$ along
the $(p$+$1)$-th axis, can be obtained by boosting the one for
parallel branes (see~\rf{backgr}) along the $(p$+$1)$-th
axis:
\begin{eqnarray}
A^{\rm cl}_0 & = & \left ( \begin{array}{cc} B_0 \cosh \epsilon & 0 \\
                0 & B_0 \cosh \epsilon \end{array} \right ), \non
A^{\rm cl}_a & = & \left ( \begin{array}{cc} B_a & 0 \\
                0 & B_a \end{array} \right ), ~~~ a=1,\ldots p, \non
A^{\rm cl}_{p+1} & = & \left ( \begin{array}{cc} B_0 \sinh \epsilon & 0 \\
                0 & -B_0 \sinh \epsilon \end{array} \right ), \non
A^{\rm cl}_{p+2} & = & \left ( \begin{array}{cc} \frac{b}{2} & 0 \\
                0 & -\frac{b}{2} \end{array} \right ), \non
A^{\rm cl}_i & = & 0, ~~~ i=p+3, \ldots 9\,.
\label{backgrv}
\end{eqnarray}
Here
\begin{equation}
v = \tanh \epsilon\,,
\end{equation}
and we have denoted
\be
B_0\equiv Q_1\,,~~B_1 \equiv P_1\,,~~\ldots~~
B_{p-1}\equiv Q_{\frac{p+1}{2}}\,,~~
B_{p} \equiv P_{\frac{p+1}{2}}\,.
\ee

The substitution of~\rf{backgrv} into the one-loop (euclidean) effective
action~\rf{seff} yields~\cite{FMOSZ97}
\begin{equation}
 W=-n^{\frac{2p}{p+1}}
 \prod_{a\ne 1}L_a^{-1} \int_0^{\infty}
        \frac{ds}{s}\left ( \frac{\pi}{2s} \right )^{\frac{p}{2}}
       e^{-b^2s} \frac{(\cosh(4\omega_1s\sinh\epsilon) -
        4\cosh(2\omega_1s\sinh\epsilon) + 3)}
        {\cosh\epsilon~\sinh(2\omega_1s\sinh\epsilon)}\,.
\label{otvetv}
\end{equation}

Using Eqs.~\rf{omega}, \rf{volume} and Wick rotating back to
Minkowski space-time, we get for the phase shift
\begin{equation}
\delta=
-\frac{V_p}{(2\pi)^p}\omega_1\prod_{i=1}^{l}\frac{1}{\omega_i^2}
        \int_0^{\infty}
        \frac{ds}{s}\left ( \frac{\pi}{2s} \right )^{\frac{p}{2}}
        e^{-b^2s}\frac{(\cos(4\omega_1s\sinh\epsilon) -
        4\cos(2\omega_1s\sinh\epsilon) + 3)}
        {\cosh\epsilon~\sin(2\omega_1s\sinh\epsilon)}
\label{wicky}
\end{equation}
where
\begin{equation}
V_p=\prod_{a=1}^{p}L_a\,.
\end{equation}

This result was shown~\cite{FMOSZ97} to agree with the superstring
calculation of Bachas~\cite{Bachas} at large $b$ for the real part
of $\delta$. Analogously, the imaginary part of~\rf{wicky} which
comes from the poles at zeros of the denominator agrees
at small $v$ providing $\omega_i=2\pi \alpha^\prime$.

\subsection{Rotated branes}

Taking the configuration of two parallel Dp-branes separated
by the distance $b$ along
the $(p$+$2)$-th axis and rotating them
in the opposite directions in the $(p,p$+$1)$ plane
through the angle $\theta /2$, one obtains the following
solution to \eq{ce}
\begin{eqnarray}\label{backgrrp}
 A_{a}^\cl&=&\left(
 \begin{array}{cc}
 B_a & 0 \\
 0 & B_a \\
 \end{array}
 \right),~~~~a=0,\ldots,p-1,\non
 A_{p}^\cl&=&\left(
 \begin{array}{cc}
 B_p\cos\frac{\theta }{2} & 0 \\
 0 & B_p\cos\frac{\theta }{2} \\
 \end{array}
 \right),      \non
 A_{p+1}^\cl&=&\left(
 \begin{array}{cc}
 B_p\sin\frac{\theta }{2} & 0 \\
 0 & -B_p\sin\frac{\theta }{2} \\
 \end{array}
 \right),          \non
 A_{p+2}^\cl&=&\left(
 \begin{array}{cc}
 \frac{b}{2} & 0 \\
 0 & -\frac{b}{2} \\
 \end{array}
 \right),\non
 A_{i }^\cl&=&0,~~~~i =p+3,\ldots,9 \,,
 \end{eqnarray}
 which extends~\rf{backgrr} to $p>1$.
This looks pretty much like an analytic continuation of \eq{backgrv}
($\epsilon\ra i\theta/2$).

The interaction between two rotated Dp-branes is given by~\cite{FMOSZ97}
\begin{eqnarray}\label{otvetr}
 W&=&-4n^{\frac{2p}{p+1}}
 \frac{1}{\cos\frac{\theta}{2}}\prod_{a\ne p-1}L_a^{-1}
 \non&&\times
 \int\limits_{0}^{\infty }\frac{ds}{s}
 \left(\frac{\pi}{2s}\right)^{\frac{p}{2}}\,
 \e^{-b^2s}\tanh\left(\omega _{\frac{p+1}{2}}s\sin\frac{\theta }{2}\right)
 \sinh^2\left(\omega _{\frac{p+1}{2}}s\sin\frac{\theta }{2}\right).
 \end{eqnarray}
It can be obtained from~\rf{otvetv} substituting $\epsilon=i\theta/2$.

Expanding in $1/b^2$ and using \eq{omega}, one gets
 \begin{equation}\label{answ}
 W=-\frac{1}{16}\,n\left(\frac{4-p}{2}\right)!\,\frac{L_{p-1}}{\sqrt{2\pi
 }n^{\frac{2}{p+1}}}\,\omega^3_{\frac{p+1}{2}}\prod_{i}
 \omega _i^{-1}
 \tan\frac{\theta }{2}\,\sin^2\frac{\theta }{2}\,\,\frac{1}{b^{6-p}}
 +O\left(\frac{1}{b^{8-p}}\right)
 \end{equation}
 for large distances, which  agrees with
 the supergravity result.
 For $p=1$ this is first shown in~\cite{ikkt}.

\subsection{The NBI action}

In the IKKT model the matrix size $n$ is considered as a
dynamical variable, so the partition function~\rf{main} includes
the summation over $n$. This sum is expected to recover the
integration over $\sqrt{g}$ in~\rf{Smain} while the proof
is missing. Even at the classical level, the minimization of \eq{action}
with respect to $n$ does not result in a nice matrix-model action which
could be associated with the Nambu--Goto action~\rf{NGaction}.

These problems can be easily resolved by a slight
modification of the IKKT matrix model. Let us
introduce a positive definite $N\times N$ hermitian matrix $Y^{ab}$
which would play the role of a dynamical variable instead of $n$.
In other words, the matrix size $N$ is fixed (to be distinguished
from fluctuating $n$) while the elements of $Y$ {\em fluctuate}.

The classical action has the form
\beq
S^{\rm cl}=-\alpha\left( \frac{1}{4}\tr Y^{-1}[A_{\mu},A_{\nu}]^2
            +\frac{1}{2}\tr (\bar{\psi}
           \g^{\mu}[A_{\mu},\psi])\right)+\beta \tr Y\,,
\label{Yaction}
\ee
which yields
the following classical equation of motion for the $Y$-field:
\be
\frac{\alpha}{4}
\left( Y^{-1}[A_{\mu},A_{\nu}]^2 Y^{-1}  \right)_{ij} +
\beta\delta_{ij}=0 \,.
\label{ceY}
\ee

The solution to \eq{ceY} reads
\be
Y= \frac{1}{2}\sqrt{\frac\alpha\beta}\sqrt{ -[A_{\mu},A_{\nu}]^2 } \,.
\label{Ycl}
\ee
Here $-[A_{\mu},A_{\nu}]^2$ is positive definite, since the commutator is
anti-hermitian (cf.\ \eq{fmunu}). The square root in~\rf{Ycl}
is unique, provided $Y$ is positive definite which is the case.
After the substitution of~\rf{Ycl}, the classical action~\rf{Yaction}
reduces to
\beq
S_{\rm NBI}^{\rm cl}=\sqrt{\alpha\beta} \tr \sqrt{-[A_{\mu},A_{\nu}]^2}
            -\frac{\alpha}{2}\tr (\bar{\psi}
           \g^{\mu}[A_{\mu},\psi])\,.
\label{NBIaction}
\ee

The bosonic part of~\rf{NBIaction}
coincides with the strong field limit of
the Non-abelian Born--Infeld (NBI) action. The action~\rf{NBIaction}
is called for this reason the NBI action. Notice that it is
{\em field-theoretic}\/ rather than widely discussed
stringy NBI action which has a different
structure~\cite{Tse97}.

The formulas above in this subsection are very similar to
the ones of Subsect.~\ref{Schi} for the Schild formulation.
Thus the hermitian matrix $Y^{ab}$ with positive definite eigenvalues
is the direct analog of $\sqrt{g(\sigma_1,\sigma_2)}$ so that
\be
 Y^{ab} \Longrightarrow  \sqrt{g(\sigma_1,\sigma_2)}
\label{r3}
\ee
in the same sense as in~\rf{r1}, \rf{r2}.
In the next subsection we discuss that it is possible to choose such
a measure
of integration over $Y$ which reproduces the Nambu--Goto version of the
Green--Schwarz action even at the quantum level.

\subsection{The NBI model of IIB superstring}

The NBI matrix model is defined by the action
\begin{equation}
S_{\rm NBI}=-\alpha \left(\frac{1}{4}\tr  Y^{-1} [A_\mu,A_\nu]^2
+\frac{1}{2} \tr \left(\bar{\psi}\g^\mu[A_\mu,\psi]
\right)\right) +V(Y)\,,
\label{SNBI}
\end{equation}
where $Y$ is a hermitian $N\times N$ matrix with {\em positive}\/
eigenvalues. The potential is
\begin{equation}
V(Y)=\beta~{\rm Tr}~Y+\gamma~{\rm Tr}~\ln Y\,,
\label{V}
\end{equation}
where
\be
\gamma =N-\frac12 \,.
\label{gamma}
\ee

The partition function is then given by the matrix integral~\cite{FMOSZ97}
\begin{equation}
Z_{\rm NBI}=\int dA_\mu \,d \psi_\alpha\,d Y\,\e^{-S_{\rm NBI}}.
\label{ZNBI}
\end{equation}
The action~\rf{SNBI} is invariant under the SUSY transformation
\begin{eqnarray}
\delta_{\rm SUSY}\:\psi&=&\frac{i}{4}\left[ Y^{-1},
\left[A_\mu,A_\nu\right]\right]_+\,
\g^{\mu\nu}\epsilon +\xi \non
\delta_{\rm SUSY} \, A_\mu&=&i\bar{\epsilon}\,\g_\mu\psi,
\label{p4}
\end{eqnarray}
in the limit $N\rightarrow\infty$, where $\left[\cdot,\cdot\right]_+$
stands for the anticommutator.
$Y$ is not changed under this transformation.

The action~\rf{SNBI} differs from its classical counterpart~\rf{Yaction}
by the second term on the right-hand side of \eq{V}. It is associated
with the measure of integration over $Y$ rather
than with the classical action. The classical action~\rf{Yaction}
can be obtained from~\rf{SNBI} in the limit $\alpha\sim\beta\ra\infty$,
$\alpha/\beta \sim 1 \sim \gamma/N$. This corresponds to
the usual classical limit in string theory since~\cite{ikkt}
$\alpha\sim\beta\sim g_s^{-1}$.

The matrix $Y$ can be always brought to the diagonal form
\be
Y=\Omega^\dagger\, {\rm diag}\, \left(y_1,\ldots,y_N \right)
\Omega~~~~~(y_1,\ldots,y_N\geq0 )\,,
\ee
where $\Omega$ is unitary. The measure for integration over $Y$
reads explicitly
\be
\int dY\,\ldots = \int_0^\infty \prod_{i=1}^N dy_i\, \Delta^2[y]
\cdot d\Omega \, \ldots
\ee
with
\be
\Delta[y]=\prod_{i>j} \left( y_i-y_j\right)
\ee
being the Vandermonde determinant.

The integral over $Y$ in~\rf{ZNBI} can be done.
Let us mention that the fermionic term in~\rf{SNBI}
is $Y$-independent and denote
\begin{equation}
{\cal F}(z)=\int dY\, \e^{ -\alpha \tr
Y^{-1} z^2/4 -\beta \tr Y-\gamma \tr\ln Y},
\label{p5}
\end{equation}
where $z^2=-[A_\mu,A_\nu]^2$. This matrix integral looks like
an external field problem for the Penner matrix model.


Doing the Itzykson--Zuber integral over the ``angular'' variable $\Omega$,
\rf{p5} takes the form
\begin{equation}
{\cal F}(z) \propto
\int_0^\infty \prod_{i=1}^{N}dy_i\,\frac{\Delta^2[y]}{\Delta[1/y]
\Delta[z^2]}\e^{-\alpha\sum_i y_i^{-1}z_i^2/4-\beta\sum_i y_i-\gamma
\sum_i\ln y_i} \non
\propto \frac{\Delta[z]}{\Delta[z^2]}
\e^{-\sqrt{\alpha\beta}\sum_i z_i},
\label{z}
\end{equation}
where $z_i^2$ stand for the eigenvalues of $z^2$.

Hence, it is shown that
\be
\int dA_\mu \,d \psi_\alpha\,d Y\,\e^{-S_{\rm NBI}} =
\int \frac{dA_\mu \,d \psi_\alpha}{\prod_{i>j}(z_i+z_j)}
\e^{-S^{\rm cl}_{\rm NBI} }.
\label{mmeasure}
\ee
Thus the NBI action $S^{\rm cl}_{\rm NBI}$ defined by \eq{NBIaction}
is reproduced modulo the change of the measure
for integration over $A_\mu$.

The significance of this result is that it can be explicitly shown
that
$$
S^{\rm cl}_{\rm NBI} \Longrightarrow S_{\rm NG}
$$
given by \eq{NGaction}, where the arrow is in the same sense as
in~\rf{r1}, \rf{r2} and \rf{r3}. Analogously, the Schild
action~\rf{Saction} can be reproduced from the model~\rf{ZNBI} with
the additional integration over $Y$ (without explicitly doing it).

A proposal of Ref.~\cite{FMOSZ97} is to modify the measure
for the integration over $A_\mu$ from
the outset to get
\bea
\lefteqn{
\int dA_\mu \,d \psi_\alpha\,d Y\,{\prod_{i>j}(z_i+z_j)}\,\e^{-S_{\rm NBI}}
=\int dA_\mu \,d \psi_\alpha
\e^{-S^{\rm cl}_{\rm NBI} }  }\non
& &\stackrel{N=\infty}{=}
\int DX^\mu \,D \Psi_\alpha
\e^{-S_{\rm NG }}. ~~~~~~~~~~~~~~~~~~~~~~~~~~~~~~~~~~~~
\label{measure}
\eea
Then the Nambu--Goto version of the Green--Schwartz action of
IIB superstring is exactly reproduced by the NBI matrix model.

\subsection{Remark on D-brane solutions in the NBI model}

The classical
 solutions~\rf{Dp} associated with D-brane configurations
 are also classical solutions to the NBI matrix model
 whose classical equations of motion, which result from the variation
 of the action~\rf{NBIaction} with respect to $A_\mu$ and $\psi_\alpha$,
 read
\be
\left[ A^{\mu},\left[Y^{-1},
\left[ A_\mu,A_\nu \right]\right]_+\right]=0\,,~~~~~
\left[ A_\mu\,, (\g^\mu\psi)_\alpha \right] =0\,.
\label{ceYA}
\ee
 The reason is that these classical solutions are BPS states
 and the commutator $[A_\mu,A_\nu]$ is proportional
 to the unit matrix (see \eq{BPS}).

 A more general property holds in the large--$N$ limit
when any classical solution of the IKKT matrix model is simultaneously
a solution of the classical equations of motion of the NBI model.
However, the structure of the classical equations~\rf{ceY} and
\rf{ceYA} in the NBI matrix model is, generally speaking,
richer that \eq{ce} in the IKKT model, since $Y^{\rm cl}$ may
have some nontrivial distribution of eigenvalues (typical
for the large--$N$ saddle points).

One of most urgent checks of the NBI model would be to perform the
calculation of the brane-brane interaction to compare with
the superstring result. This calculation will take into account
the fact that $Y$ is a dynamical field while the ones described
above for the IKKT matrix model are done at fixed $n$, \ie without
considering $n$ as a dynamical variable.

\eop
\section*{Conclusion}

It is now too early to make any definite conclusions since it is
not yet clear whether or not this formulation of superstrings, which
is based on the supersymmetric matrix models, would survive.
Nevertheless, such an approach to M~theory looks most promising
among those proposed so far.

This situation reminds me somewhat of the one with QCD in the very
beginning of the seventies about the time when the QCD Lagrangian
was introduced. Before that there existed the approach to
the theory of strong interaction based on strings
and dual resonance models, while the
new theory looked quite different and was most convenient
to study strong interaction at small distances.
Once again, it is now too early to predict whether the same could happen
with superstrings in the nearest future, but this option should not
be immediately excluded.

One of the simplest checks of the matrix models of superstrings
is the study of the interaction between D-branes.
It should answer, in particular, the question whether
the classical operator-like solutions of the matrix models are
associated with empty D-branes or D-branes carrying magnetic field.

A more serious problem is to show how
string perturbation theory emerges from the matrix models.
The NBI matrix model is very promising from this point of view since it
reproduces the Nambu--Goto version of the Green--Schwartz action.

While the proposed matrix models of IIB superstring are of the type
of reduced ten-dimensional super Yang--Mills, they have additional
degrees of freedom which are essential to have strings.
This differs the situation from the one in large--$N$ QCD where
the fundamental Lagrangian is fixed, and the problem to
obtain strings in the Eguchi--Kawai reduced model is almost
as difficult as in whole QCD. Now, for the matrix models of
superstrings, the true model is not know from the outset.
The reader is still free to introduce his/her own
model to describe superstrings in the best way

\subsection*{Acknowledgments}

I thank I.~Chepelev, A.~Fayyazuddin, P.~Olesen, D.~Smith  and K.~Zarembo
for the pleasant collaboration and illuminating discussions,
and K.~Montonen for the hospitality in Helsinki.
This work was supported in part by the grants INTAS 94--0840,
CRDF 96--RP1--253 and RFFI 97--02--17927.

\eop


\begin{thebibliography}{99}
\addtolength{\itemsep}{-6pt}

\bibitem{BFSS96} T. Banks, W. Fischler, S.H. Shenker and L. Susskind,
{\it M theory as a matrix model: a conjecture}, hep-th/9610043.

\bibitem{ikkt}
N.~Ishibashi, H.~Kawai, Y.~Kitazawa and A.~Tsuchiya,
{\it A large--N reduced model as superstring}, hep-th/9612115.

\bibitem{FMOSZ97}
A. Fayyazuddin, Y. Makeenko, P. Olesen, D.J. Smith and K. Zarembo,
{\it Towards a non-perturbative formulation of IIB Superstring
by Matrix Models}, hep-th/9703038.

\bibitem{Kaz85}
V.A.~Kazakov,
{\it Bilocal regularization of models of random surfaces},
Phys.\ Lett. {\bf 150B} (1985) 282.

\bibitem{Dav85}
F.~David,
{\it Planar diagrams, two-dimensional lattice gravity and surface models},
Nucl.\ Phys. {\bf B257}[FS14] (1985) 45.

\bibitem{ADF85}
J.~Ambj{\o}rn, B.~Durhuus and J.~Fr{\"o}lich,
{\it Diseases of triangulated random surfaces, and possible cures},
Nucl.\ Phys. {\bf B257}[FS14] (1985) 433.

\bibitem{Mak96}
Y.M. Makeenko, {\it Discretizing Superstring},
Talk at QUARKS'96, Yaroslavl May 5--11, 1996,
ITEP--TH--42/96 (September, 1996).

\bibitem{MS90}
A. Mikovi{\'c} and W. Siegel,
{\it Random superstrings},
Phys.\ Lett. {\bf 240B} (1990) 363.

\bibitem{AG92}
L. Alvarez-Gaume {\it et al.},
{\it Superloop equations and two dimensional supergravity},
{Int.\ J.\ Mod.\ Phys.} {\bf A7} (1992) 5337.

\bibitem{Pol95}
J.~Polchinski,
{\it Dirichlet branes and Ramond--Ramond charges},
Phys.\ Rev.\ Lett. {\bf 75} (1995) 4724.

\bibitem{Ste97}
K.S. Stelle,
{\it Lectures on supergravity p-branes},
hep-th/9701088.

\bibitem{Wit95}
E. Witten,
{\it Bound states of strings and $p$-branes},
Nucl.\ Phys. {\bf B460} (1995) 335.

\bibitem{dWHN88}
B. de Wit, J. Hoppe and H. Nicolai,
{\it On the quantum mechanics of supermembranes},
Nucl.\ Phys. {\bf B305}[FS23] (1988) 545.

\bibitem{dWLN89}
B. de Wit, M. Luscher and H. Nicolai,
{\it The supermembrane is unstable},
Nucl.\ Phys. {\bf B320} (1989) 135.

\bibitem{AB96}
O. Aharony and M. Berkooz,
{\it Membrane Dynamics in M(atrix) Theory},
hep-th/9611215.

\bibitem{LM96}
G. Lifschytz and S.D. Mathur,
{\it Supersymmetry and Membrane Interactions in M(atrix) Theory},
hep-th/9612087.

\bibitem{Lif96}
G. Lifschytz,
{\it Four-Brane and Six-Brane Interactions in M(atrix)Theory},
hep-th/9612223.

\bibitem{bss}
T.~Banks, N.~Seiberg and S.~Shenker,
{\it Branes from Matrices}, hep-th/9612157.

\bibitem{CMZ97}
I.~Chepelev, Y.~Makeenko and K.~Zarembo,
{\it Properties of D-Branes in Matrix Model of IIB Superstring},
hep-th/9701151.

\bibitem{FS97}
A.~Fayyazuddin and D.J.~Smith,
{\it P-Brane Solutions in IKKT IIB Matrix Theory}, hep-th/9701168.

\bibitem{Pol96}
J.~Polchinski, {\it TASI Lectures on D-Branes}, hep-th/9611050.

\bibitem{GG96}
M.B. Green and M. Gutperle,
{\it Light-cone supersymmetry and D-branes},
Nucl. Phys. {\bf B476} (1996) 484.

\bibitem{Tse97}
A.A. Tseytlin,
{\it On non-Abelian Generalization of Born--Infeld Action
in String Theory}, hep-th/9701125.

\bibitem{Bachas}
C. Bachas,
{\it D-brane dynamics},
Phys.\ Lett. {\bf B374} (1996) 37.


\end{thebibliography}
\end{document}